\documentclass{ws-procs961x669}            
\usepackage{array}
\newcommand{\be}{\begin{eqnarray}}
\newcommand{\ee}{\end{eqnarray}}
\usepackage[colorlinks]{hyperref}
\usepackage{amsmath,amssymb}
\usepackage{mathtools}

\begin{document}
\title{Probing Gluon Sivers Function in Inelastic Photoproduction of
$J/\psi$ at the EIC}

\author{Raj Kishore and Asmita Mukherjee$^*$}

\address{Department of Physics, Indian Institute of Technology Bombay,
Powai, Mumbai 4000076, India\\
$^*$E-mail: asmita@phy.iitb.ac.in}

\author{Sangem Rajesh}

\address{Dipartimento di Fisica, Universit\`a di Cagliari, Cittadella Universitaria, 
I-09042 Monserrato (CA), Italy
and INFN, Sezione di Cagliari, Cittadella Universitaria, I-09042 Monserrato (CA), Italy}

\begin{abstract}
We present a recent calculation of the single spin asymmetry in inclusive
photoproducton of $J/\psi$ at the future EIC, that can be used to probe the
gluon Sivers function.
\end{abstract}


\bodymatter

\section{Introduction}\label{intro}
$J/\psi$ production in $ep$ and $pp$ collisions is known to be an effective
tool to probe the gluon TMDs, as the contribution comes at leading order
(LO) through $\gamma g$ and $gg$ initiated processes. One of the most
interesting gluon TMDs is the gluon Sivers function (GSF) which probes the
coupling of the intrinsic transverse momenta of the gluons with the
transverse spin of the nucleon. The Sivers function \cite{Sivers:1989cc,
Sivers:1990fh} is a time reversal odd
(T-odd) object and initial and final state interactions play an important
role in the Sivers asymmetry. The gluon Sivers function for
any process can be written as a linear combination of two gluon Sivers
functions, one containing a C-even operator (f-type)  and the other
C-odd (d-type) \cite{Buffing:2013kca}. Compared to the quark Sivers function, much less is known
about the gluon Sivers function, apart from a positivity bound
\cite{Mulders:2000sh}. In this talk,
we present a recent calculation \cite{Rajesh:2018qks}
of single spin asymmetry in inelastic photoproduction of $J/\psi$ at the
future EIC.

\section{Calculation of the asymmetry}     

The process considered is
\be
e(l)+p^\uparrow(P) \rightarrow J/\psi (P_h) +X 
\ee
where the quantities within brackets are the momenta. We use the kinematics where the interaction takes place 
through the exchange of an (almost) real photon $\gamma(q)+ g(k) \rightarrow J/\psi(P_h) +g(p_g)$.  
We consider only the direct photon contribution and contribution from the resolved photon is eliminated by imposing a cut on the variable $z= {P \cdot P_h \over P\cdot q}$, which is the energy fraction transferred from the photon to the $J/\psi$ in the rest frame of the proton. 
The inelasticity variable  $z$ for inclusive photoproduction can be measured in experiments by Jacquet-Blondel method.  
The leading order (LO) process $ \gamma+g \rightarrow J/\psi$ contributes at $z=1$ (see
\cite{Mukherjee:2016qxa})
for the calculation of Sivers asymmetry in electroproduction) and we
use a cutoff $z<0.9$ to remove this contribution. Also contribution to
$J/\psi$ production from gluon and heavy quark fragmentation was removed  
by imposing a cut on $P_T$, which is the transverse momentum of the
$J/\psi$. We assume TMD factorization for the process considered and
generalized parton model (GPM) with the inclusion of the intrinsic transverse
momenta. We use NRQCD \cite{Bodwin:1994jh} to calculate the production of $J/\psi$. The $c{\bar
c}$ pair can be produced in color singlet (CS) or color octet (CO) state. In
$eP$ collision, non-zero asymmetry can be observed only if the $c {\bar c}$
pair is produced in the CO state \cite{Yuan:2008vn}. The SSA is defined as            

\begin{eqnarray}\label{asy}
A_N=\frac{d \sigma^{\uparrow}-d 
\sigma^{\downarrow}}{d \sigma^{\uparrow}+d \sigma^{\downarrow}},
\end{eqnarray}

where $d\sigma^{\uparrow}$ and $d\sigma^{\downarrow}$ are the  differential cross-sections 
measured when one of the particle is transversely polarized up ($\uparrow$) and down
($\downarrow$), respectively,  with respect to the scattering plane.

We consider the inclusive process   $e(l)+p^{\uparrow}(P)\rightarrow 
J/\psi(P_h)+X$. The virtual photon radiated by the initial electron is
almost real, $q^2=-Q^2 \approx 0$. The numerator and the denominator of the
asymmetry are given by 
\be\label{d3}
 \begin{aligned}
 d\sigma^{\uparrow}-d\sigma^{\downarrow}={}&\frac{d\sigma^{ep^{\uparrow}\rightarrow J/\psi X}}{dzd^2{\bm 
P}_T}-\frac{d\sigma^{ep^{\downarrow}\rightarrow J/\psi X}}{dzd^2{\bm P}_T}\\
={}&\frac{1}{2z(2\pi)^2}\int dx_\gamma dx_g  
d^2{\bm k}_{\perp g}
f_{\gamma/e}(x_\gamma)\Delta^N f_{g/p^{\uparrow}}(x_g,{\bm k}_{\perp 
g})\\
&\times\delta(\hat{s}+\hat{t}+\hat{u}-M^2)\frac{1}{2\hat{s}}|\mathcal{M}_{\gamma+g\rightarrow J/\psi +g}|^2,
\end{aligned}
\ee
and 
\be\label{d4}
\begin{aligned}
d\sigma^{\uparrow}+d\sigma^{\downarrow}={}&\frac{d\sigma^{ep^{\uparrow}\rightarrow  J/\psi X}}{dzd^2{\bm 
P}_T}+\frac{d\sigma^{ep^{\downarrow}\rightarrow  J/\psi X}}{dzd^2{\bm P}_T}=2\frac{d\sigma}{dzd^2{\bm P}_T}\\
={}&\frac{2}{2z(2\pi)^2}\int dx_{\gamma} dx_g 
d^2{\bm k}_{\perp g}
f_{\gamma/e}(x_\gamma) f_{g/p}(x_g,{\bm k}_{\perp 
g})\\
&\times\delta(\hat{s}+\hat{t}+\hat{u}-M^2)\frac{1}{2\hat{s}}|\mathcal{M}_{\gamma+g\rightarrow J/\psi +g}|^2.
\end{aligned}
\ee 

$x_\gamma$ and $x_g$ are the light-cone momentum fractions of the photon and gluon respectively. 
The Weizs$\ddot{a}$ker-Williams distribution function, $f_{\gamma/e}(x_\gamma)$, describes the density of photons inside the
electron. The $J/\psi $ production rate is calculated in NRQCD based color
octet model. 
We follow the approach given in \cite{Boer:2012bt} and the details of the calculation can be
found in \cite{Rajesh:2018qks}.  Here, we report on some of our numerical results in the
kinematics of the future planned EIC.   
\section{Results}
For the numerical estimates of the SSA we assume Gaussian parametrization of
unpolarized TMDs and best fit parameters from \cite{alesio} for the gluon Sivers
function. These are denoted by SIDIS1 and SIDIS2, respectively. Also,
following \cite{Boer:2003tx}  we parametrize the GSF in terms of $u$ and $d$ quark Sivers
functions \cite{Anselmino:2016uie}. Two different choices in this line are labeled as BV-a and
BV-b. For the dominating channel of photon-gluon fusion, contribution to the
numerator of the SSA comes mainly from GSF. As the heavy quark pair is
produced unpolarized, there is no contribution from the Collins function. 
Figs 1 and 2 show plots of the SSA for $\sqrt{s}= 100$ ~GeV and $45$ ~ GeV
respectively which will be possible at the future EIC. The asymmetry depends
strongly on the parametrization for the GSF, it is positive for SIDIS1 and
SIDIS2 and negative for BV-a and BV-b parametrizations. The magnitude of the
asymmetry is largest for BV-b. We have incorporated contributions from the 
 ${^{3}}{S}{_1}^{(8)}$, ${^{1}}{S}{_0}^{(8)}$ and 
${^{3}}{P}{_{J(0,1,2)}}^{(8)}$ in the asymmetry.
Numerical estimates of the unpolarized cross
section shows that the data from HERA can be explained if both CS and CO
contributions are incorporated.

\begin{figure}[]
\begin{minipage}[c]{0.99\textwidth}
\small{(a)}\includegraphics[width=5.5cm,height=5.5cm,clip]{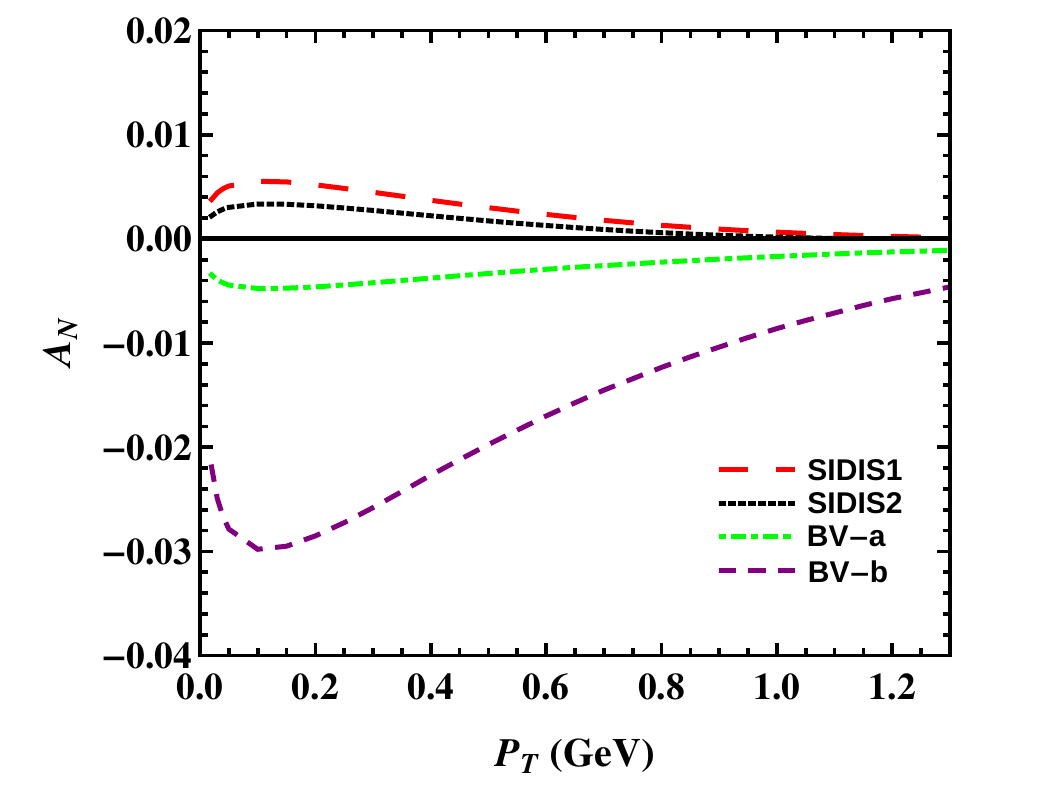}
\hspace{0.1cm}
\small{(b)}\includegraphics[width=5.5cm,height=5.5cm,clip]{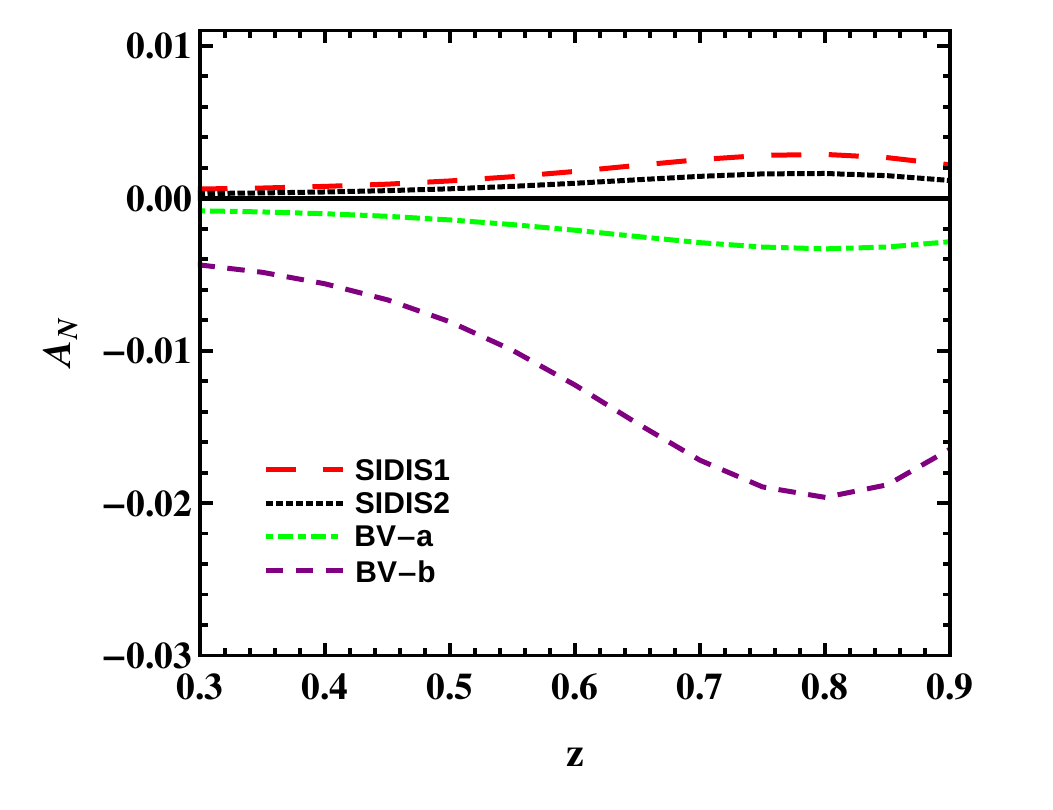}
\end{minipage}
\caption{\label{fig3}Single spin asymmetry  in $e+p^{\uparrow}\rightarrow J/\psi +X$
process as function of 
(a) $P_T$ (left panel) and  (b) $z$ (right panel) at $\sqrt{s}=100$ GeV (EIC)
\cite{Rajesh:2018qks}. The integration ranges are 
$0<P_{T}\leq1$ GeV and $0.3<z<0.9$. For convention of lines see the text.}
\end{figure}
\begin{figure}[]
\begin{minipage}[c]{0.99\textwidth}
\small{(a)}\includegraphics[width=5.5cm,height=5.5cm,clip]{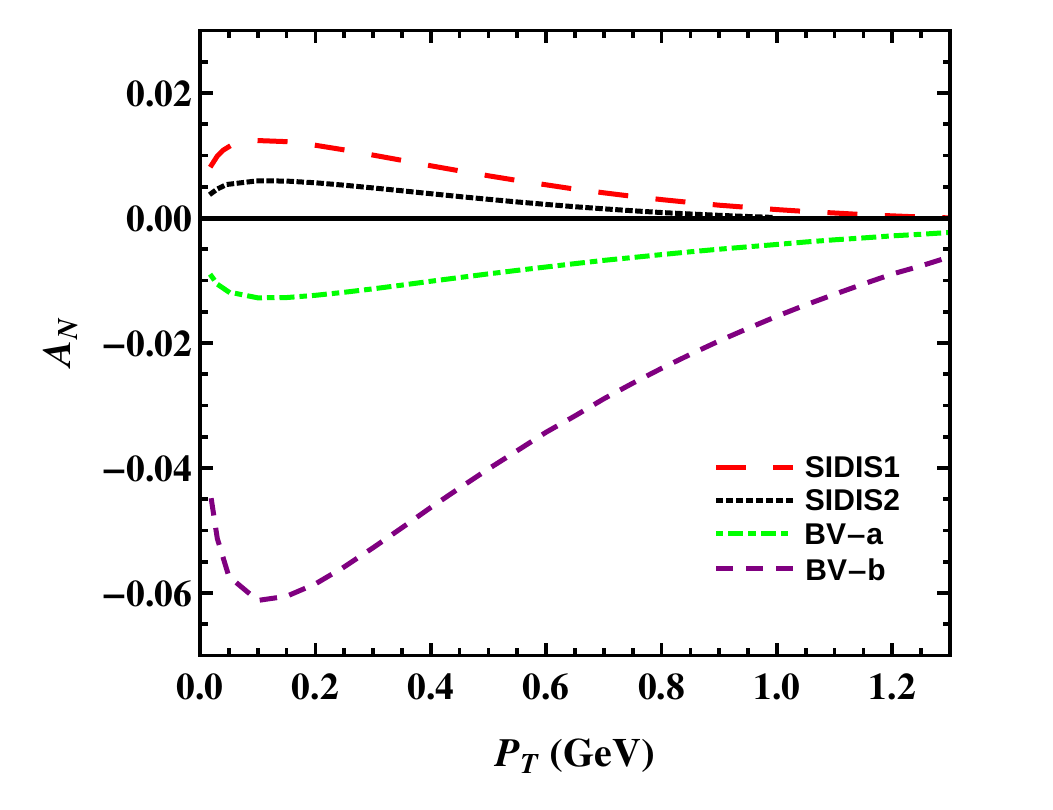}
\hspace{0.1cm}
\small{(b)}\includegraphics[width=5.5cm,height=5.5cm,clip]{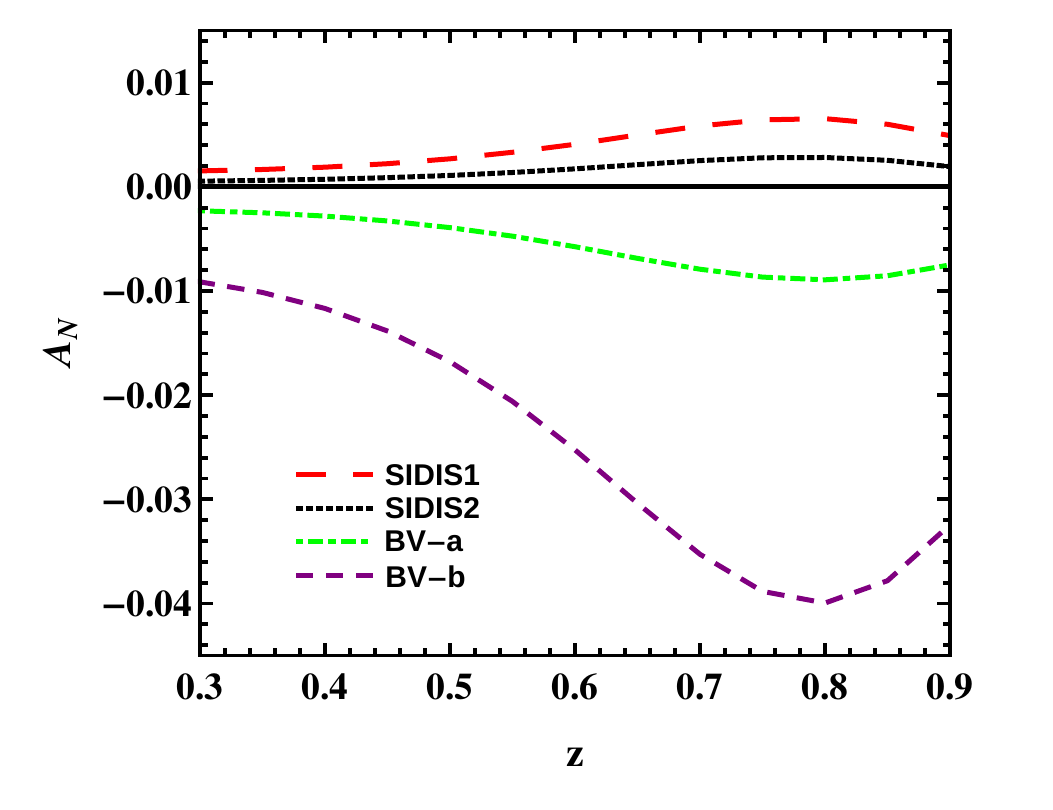}
\end{minipage}
\caption{\label{fig3}Single spin asymmetry  in $e+p^{\uparrow}\rightarrow J/\psi +X$
process as function of 
(a) $P_T$ (left panel) and  (b) $z$ (right panel) at $\sqrt{s}=45$ GeV (EIC). The 
integration ranges are  $0<P_{T}\leq1$ GeV and $0.3<z<0.9$
\cite{Rajesh:2018qks}. For convention of lines see the text.}
\end{figure}

\section{Conclusion}

We have presented a recent calculation of the SSA in inclusive
photoproduction of $J/\psi$ production in the kinematics of the future EIC. 
A sizable asymmetry in NRQCD based color octet model is reported, for the
range $0 < P_T < 1 $ ~ GeV and $0.3 < z < 0.9$. This asymmetry can give
direct access to the GSF.           

\section{Acknowledgement}
AM would like to thank the organizers of INT program "Probing Nucleons and
Nuclei in High Energy Collisions", October 1- November 16, 2018 for the
invitation.  

\bibliographystyle{apsrev}
\bibliography{refer}

\begin{thebibliography}{12}
\expandafter\ifx\csname natexlab\endcsname\relax\def\natexlab#1{#1}\fi
\expandafter\ifx\csname bibnamefont\endcsname\relax
  \def\bibnamefont#1{#1}\fi
\expandafter\ifx\csname bibfnamefont\endcsname\relax
  \def\bibfnamefont#1{#1}\fi
\expandafter\ifx\csname citenamefont\endcsname\relax
  \def\citenamefont#1{#1}\fi
\expandafter\ifx\csname url\endcsname\relax
  \def\url#1{\texttt{#1}}\fi
\expandafter\ifx\csname urlprefix\endcsname\relax\def\urlprefix{URL }\fi
\providecommand{\bibinfo}[2]{#2}
\providecommand{\eprint}[2][]{\url{#2}}

\bibitem[{\citenamefont{Sivers}(1990)}]{Sivers:1989cc}
\bibinfo{author}{\bibfnamefont{D.~W.} \bibnamefont{Sivers}},
  \bibinfo{journal}{Phys. Rev.} \textbf{\bibinfo{volume}{D41}},
  \bibinfo{pages}{83} (\bibinfo{year}{1990}).

\bibitem[{\citenamefont{Sivers}(1991)}]{Sivers:1990fh}
\bibinfo{author}{\bibfnamefont{D.~W.} \bibnamefont{Sivers}},
  \bibinfo{journal}{Phys. Rev.} \textbf{\bibinfo{volume}{D43}},
  \bibinfo{pages}{261} (\bibinfo{year}{1991}).

\bibitem[{\citenamefont{Buffing et~al.}(2013)\citenamefont{Buffing, Mukherjee,
  and Mulders}}]{Buffing:2013kca}
\bibinfo{author}{\bibfnamefont{M.~G.~A.} \bibnamefont{Buffing}},
  \bibinfo{author}{\bibfnamefont{A.}~\bibnamefont{Mukherjee}},
  \bibnamefont{and} \bibinfo{author}{\bibfnamefont{P.~J.}
  \bibnamefont{Mulders}}, \bibinfo{journal}{Phys. Rev.}
  \textbf{\bibinfo{volume}{D88}}, \bibinfo{pages}{054027}
  (\bibinfo{year}{2013}), \eprint{1306.5897}.

\bibitem[{\citenamefont{Mulders and Rodrigues}(2001)}]{Mulders:2000sh}
\bibinfo{author}{\bibfnamefont{P.~J.} \bibnamefont{Mulders}} \bibnamefont{and}
  \bibinfo{author}{\bibfnamefont{J.}~\bibnamefont{Rodrigues}},
  \bibinfo{journal}{Phys. Rev.} \textbf{\bibinfo{volume}{D63}},
  \bibinfo{pages}{094021} (\bibinfo{year}{2001}), \eprint{hep-ph/0009343}.

\bibitem[{\citenamefont{Rajesh et~al.}(2018)\citenamefont{Rajesh, Kishore, and
  Mukherjee}}]{Rajesh:2018qks}
\bibinfo{author}{\bibfnamefont{S.}~\bibnamefont{Rajesh}},
  \bibinfo{author}{\bibfnamefont{R.}~\bibnamefont{Kishore}}, \bibnamefont{and}
  \bibinfo{author}{\bibfnamefont{A.}~\bibnamefont{Mukherjee}},
  \bibinfo{journal}{Phys. Rev.} \textbf{\bibinfo{volume}{D98}},
  \bibinfo{pages}{014007} (\bibinfo{year}{2018}), \eprint{1802.10359}.

\bibitem[{\citenamefont{Mukherjee and Rajesh}(2017)}]{Mukherjee:2016qxa}
\bibinfo{author}{\bibfnamefont{A.}~\bibnamefont{Mukherjee}} \bibnamefont{and}
  \bibinfo{author}{\bibfnamefont{S.}~\bibnamefont{Rajesh}},
  \bibinfo{journal}{Eur. Phys. J.} \textbf{\bibinfo{volume}{C77}},
  \bibinfo{pages}{854} (\bibinfo{year}{2017}), \eprint{1609.05596}.

\bibitem[{\citenamefont{Bodwin et~al.}(1995)\citenamefont{Bodwin, Braaten, and
  Lepage}}]{Bodwin:1994jh}
\bibinfo{author}{\bibfnamefont{G.~T.} \bibnamefont{Bodwin}},
  \bibinfo{author}{\bibfnamefont{E.}~\bibnamefont{Braaten}}, \bibnamefont{and}
  \bibinfo{author}{\bibfnamefont{G.~P.} \bibnamefont{Lepage}},
  \bibinfo{journal}{Phys. Rev.} \textbf{\bibinfo{volume}{D51}},
  \bibinfo{pages}{1125} (\bibinfo{year}{1995}), \bibinfo{note}{[Erratum: Phys.
  Rev.D55,5853(1997)]}, \eprint{hep-ph/9407339}.

\bibitem[{\citenamefont{Yuan}(2008)}]{Yuan:2008vn}
\bibinfo{author}{\bibfnamefont{F.}~\bibnamefont{Yuan}}, \bibinfo{journal}{Phys.
  Rev.} \textbf{\bibinfo{volume}{D78}}, \bibinfo{pages}{014024}
  (\bibinfo{year}{2008}), \eprint{0801.4357}.

\bibitem[{\citenamefont{Boer and Pisano}(2012)}]{Boer:2012bt}
\bibinfo{author}{\bibfnamefont{D.}~\bibnamefont{Boer}} \bibnamefont{and}
  \bibinfo{author}{\bibfnamefont{C.}~\bibnamefont{Pisano}},
  \bibinfo{journal}{Phys. Rev.} \textbf{\bibinfo{volume}{D86}},
  \bibinfo{pages}{094007} (\bibinfo{year}{2012}), \eprint{1208.3642}.

\bibitem[{\citenamefont{D'Alesio et~al.}(2015)\citenamefont{D'Alesio, Murgia,
  and Pisano}}]{alesio}
\bibinfo{author}{\bibfnamefont{U.}~\bibnamefont{D'Alesio}},
  \bibinfo{author}{\bibfnamefont{F.}~\bibnamefont{Murgia}}, \bibnamefont{and}
  \bibinfo{author}{\bibfnamefont{C.}~\bibnamefont{Pisano}},
  \bibinfo{journal}{JHEP} \textbf{\bibinfo{volume}{09}}, \bibinfo{pages}{119}
  (\bibinfo{year}{2015}), \eprint{1506.03078}.

\bibitem[{\citenamefont{Boer and Vogelsang}(2004)}]{Boer:2003tx}
\bibinfo{author}{\bibfnamefont{D.}~\bibnamefont{Boer}} \bibnamefont{and}
  \bibinfo{author}{\bibfnamefont{W.}~\bibnamefont{Vogelsang}},
  \bibinfo{journal}{Phys. Rev.} \textbf{\bibinfo{volume}{D69}},
  \bibinfo{pages}{094025} (\bibinfo{year}{2004}), \eprint{hep-ph/0312320}.

\bibitem[{\citenamefont{Anselmino et~al.}(2017)\citenamefont{Anselmino,
  Boglione, D'Alesio, Murgia, and Prokudin}}]{Anselmino:2016uie}
\bibinfo{author}{\bibfnamefont{M.}~\bibnamefont{Anselmino}},
  \bibinfo{author}{\bibfnamefont{M.}~\bibnamefont{Boglione}},
  \bibinfo{author}{\bibfnamefont{U.}~\bibnamefont{D'Alesio}},
  \bibinfo{author}{\bibfnamefont{F.}~\bibnamefont{Murgia}}, \bibnamefont{and}
  \bibinfo{author}{\bibfnamefont{A.}~\bibnamefont{Prokudin}},
  \bibinfo{journal}{JHEP} \textbf{\bibinfo{volume}{04}}, \bibinfo{pages}{046}
  (\bibinfo{year}{2017}), \eprint{1612.06413}.

\end{thebibliography}

\end{document}